\documentstyle[12pt]{article}
\def\be{\begin{equation}}
\def\ee{\end{equation}}
\def\bea{\begin{eqnarray}}
\def\eea{\end{eqnarray}}

\begin{document}
\begin{flushright}
hep-th/9909079 \\
IPM/P-99/047
\end{flushright}

\begin{center}
{\Large{\bf Mixed Branes at Angle in Compact Spacetime }}                  
										 
\vskip .5cm   
{\large  Davoud Kamani}
\vskip .1cm
 {\it Institute for Studies in Theoretical Physics and 
Mathematics
\\ P.O.Box: 19395-5531, Tehran, Iran}\\
{\it and}\\
{\it  Department of Physics, Sharif University of Technology
P.O.Box 11365-9161,Tehran, Iran}\\
{\sl e-mail:kamani@netware2.ipm.ac.ir}
\\
\end{center}

\begin{abstract} 
In this article the interaction of branes at angles with respect to   
each other with non-zero internal gauge fields are calculated by construction
of the boundary states in spacetime in which some of its directions are
compact on tori. The interaction depends on both angle and fields.

\end{abstract} 
\vskip .5cm

\newpage
\section{Introduction}
 
A useful tool for describing branes and their interactions, specially
in non-zero back-ground fields is the boundary state \cite{1,2,3,
4,5,6,7,8}. The overlap of boundary states through the 
closed string propagator
gives us the amplitude of interaction of branes. By introducing back-ground 
fields to the string $\sigma$-model action one obtains mixed boundary
conditions (i.e. a combination of Dirichlet and Neumann boundary conditions)
for the strings. Previously we obtained the boundary state for
a stationary mixed brane (i.e. a brane in back-ground fields ). 
We also observed that when some directions are compactified
on tori \cite{6,8}, the winding numbers of the emitted states 
around the compact directions of the brane are correlated with
their momenta along the brane. Also we studied the interaction of two
of these mixed branes, in the parallel and perpendicular cases.

In this article we consider more general case of branes with respect to
one another. In particular we take two non-intersecting one-branes which
make an angle $\phi$ with each other in presence of the non-zero back-ground
$B$-field. To keep our analysis general, we keep certain directions
compact. The correct result for the non-compact case is recovered
when the radii are take to infinity.

In section 2 we obtain the boundary state for the oblique $m_1$-brane and 
interaction of two non-intersecting
angled mixed branes for the bosonic part of the
theory. In section 3 we develop these, for the NS-NS and the R-R sectors of  
superstring theory. Finally in section 4 we
extract the contribution of the massless states on the interaction of the 
branes.

We denote a brane in the back-ground  
field by ``$m_p$-brane'', which is a ``mixed brane'' with dimension ``p''.

%%%%%%%%%%%%%%%%%%%%%%%%%%%%%%%%%%%%%%%%%%%%%%%%%%%%%%%%%%%%%%%%%%%%%%%%%%%%%
\section{The bosonic part} 
{\bf Boundary state}

 Previously we obtained the boundary state of a $m_p$-brane \cite{6}.
 For the $m_1$-brane along the $X^1$-direction 
 with field strength ${\cal{F}}_{01}=E$, the boundary state 
 satisfies the following equations
 \bea
 (\partial_{\tau}X^0 - E \partial_{\sigma}X^1)_{\tau_0} \mid B_x , \tau_0
 \rangle = 0 \;\;,
 \eea
 \bea
 (\partial_{\tau}X^1 - E \partial_{\sigma}X^0)_{\tau_0} \mid B_x , \tau_0 
 \rangle = 0 \;\;,    
 \eea
 \bea
 [X^i(\sigma , \tau)-y^i]_{\tau_0} \mid B_x , \tau_0 \rangle = 0 \;\;.    
 \eea
 In these equations the set $\{y^i\}$ shows the 
 position of the $m_1$-brane with
 $i \in \{ 2,3,...,d-1 \}$. With equation (3) we have fixed the position
 of the $m_1$-brane. Also $\tau_0$ is the $\tau$ variable on the boundary
 of the closed string world sheet. Now consider a $m_1$-brane in the
 $X^1X^2$-plane, which makes angle $\theta$ with $X^1$-direction. For this
 brane, boundary state equations have the form,
 \bea
 \bigg{(} \partial_{\tau}X^0 - E \cos \theta \partial_{\sigma}X^1
 -E \sin \theta \partial_{\sigma}X^2 \bigg{)}_{\tau_0} \mid B_x , 
 \tau_0 \rangle = 0 \;\;,
 \eea
 \bea
 \bigg{(}\cos \theta \partial_{\tau}X^1 +  \sin \theta \partial_{\tau}X^2
 -E \partial_{\sigma}X^0 \bigg{)}_{\tau_0} \mid B_x , 
 \tau_0 \rangle = 0 \;\;,
 \eea
 \bea
 \bigg{(} -(X^1 - y^1) \sin \theta +(X^2-y^2)\cos \theta 
 \bigg{)}_{\tau_0} \mid B_x , \tau_0 \rangle = 0 \;\;,
 \eea
 \bea
 (X^j-y^j)_{\tau_0}\mid B_x , \tau_0 \rangle = 0 \;\;\;,\;\;\;j\neq 0,1,2 .
 \eea
 In terms of the modes $X^{\mu} (\sigma , \tau )$ has the form,
\bea
X^{\mu}(\sigma,\tau)= x^{\mu}+2{\alpha'}p^{\mu}\tau+2L^{\mu}\sigma+
 \frac{i}{2}\sqrt{2\alpha'}\sum_{m\neq 0}\frac{1}{m}\bigg{(}\alpha^{\mu}_{m}
 e^{-2im(\tau-
 \sigma)}+\tilde\alpha^{\mu}_{m}e^{-2im(\tau+\sigma)}\bigg{)} \;\; ,
\eea
where $L^{\mu}$ is zero for non-compact directions. For compact 
directions we have $L^{\mu} = N^{\mu}R^{\mu}$ and $p^{\mu} = \frac{M^{\mu}}
{R^{\mu}}$ , in which $N^{\mu}$ is the winding number and $M^{\mu}$ is the 
momentum number of the closed string state, and $R_{\mu}$ is the radius 
of compactification of $X^{\mu}$-direction. 
Suppose $X^0, X^1$ and $X^2$ to be in the set of compact
directions $\{X^{\mu_c}\}$, therefore boundary state 
equations (4)-(7) in terms of the modes have the following forms,
\bea
\bigg{(} p^0 - \frac{1}{\alpha'}E(L_1 \cos \theta + L_2 \sin \theta)
\bigg{)}_{op} \mid B_x , \tau_0 \rangle = 0 \;\;, 
\eea
\bea
(p^1 \cos \theta + p^2 \sin \theta - \frac{1}{\alpha'}EL^0 )
_{op} \mid B_x , \tau_0 \rangle = 0 \;\;,   
\eea
\bea
\bigg{(} -(x^1+2\alpha' \tau_0p^1 -y^1)\sin \theta + (x^2+2\alpha'\tau_0 p^2
-y^2)\cos \theta \bigg{)}_{op} \mid B_x , \tau_0 \rangle = 0 \;\;,
\eea
\bea
(L_1 \sin \theta - L_2 \cos \theta )_{op} \mid B_x , \tau_0 \rangle = 0 \;\;,   
\eea
\bea
(x^j + 2 \alpha' \tau_0 p^j -y^j)_{op} \mid B_x , \tau_0 \rangle = 0 \;\;,   
\eea
\bea
L^j_{op} \mid B_x , \tau_0 \rangle = 0 \;\;,   
\eea
for the zero modes, and
\bea
&~& \;\;\;\;\;\;\;\;\;
\bigg{(} [ \alpha^0_m + E(\alpha^1_m \cos \theta +\alpha^2_m \sin \theta)]
e^{-2im\tau_0}  
\nonumber\\
&~&+[ \tilde{\alpha}^0_{-m} - E(\tilde{\alpha}^1_{-m} \cos \theta 
 +\tilde{\alpha}^2_{-m} \sin \theta)] 
e^{2im\tau_0}\bigg{)}  \mid B_x , \tau_0 \rangle = 0 \;\;, 
\eea
\bea
&~&\;\;\;\;\;\;\;\;\;
\bigg{(} ( \alpha^1_m \cos \theta +\alpha^2_m \sin \theta +E \alpha^0_m)
e^{-2im\tau_0}  
\nonumber\\
&~&+( \tilde{\alpha}^1_{-m} \cos \theta 
+\tilde{\alpha}^2_{-m}\sin \theta -E\tilde{\alpha}^0_{-m})
e^{2im\tau_0}\bigg{)}  \mid B_x , \tau_0 \rangle = 0 \;\;,
\eea
\bea
\bigg{(} (-\alpha^1_m \sin \theta +\alpha^2_m \cos \theta )e^{-2im\tau_0} - 
(-\tilde{\alpha}^1_{-m} \sin \theta 
+\tilde{\alpha}^2_{-m} \cos \theta) 
e^{2im\tau_0}\bigg{)} \mid B_x , \tau_0 \rangle = 0 \;\;, 
\eea
\bea
\bigg{(} \alpha^j_m e^{-2im\tau_0} - \tilde{\alpha}^j_{-m} e^{2im\tau_0} 
\bigg{)} \mid B_x , \tau_0 \rangle = 0 \;\;, 
\eea
for the oscillatory modes. Note that $j \in \{ 3,4,...,d-1\}$. 
The oscillating parts can be written as
\bea
\bigg{(} \alpha^{\mu}_m e^{-2im\tau_0} + S^{\mu}_{\;\;\;\nu} 
\tilde{\alpha}^{\nu}_{-m} e^{2im\tau_0} 
\bigg{)} \mid B_x , \tau_0 \rangle = 0 \;\;, 
\eea
where the matrix $S$, which depends on angle $\theta$ and the electric field
$E$ is, 
\bea
S^{\mu}_{\;\;\;\nu}(\theta,E) = (\Omega^{\alpha}_{\;\;\;\beta} \;, \;
-\delta^j_{\;\;\;k} \;) \;\;\;,\;\;\;\alpha , \beta =0,1,2\;\;\;, 
\eea
\bea
\Omega^{\alpha}_{\;\;\beta} =\frac{1}{1-E^2} \left( \begin{array}{ccc}
1+E^2 & -2E \cos \theta & -2E \sin \theta \\
-2E \cos \theta & E^2+ \cos2\theta & \sin2\theta \\
-2E\sin\theta & \sin2\theta & E^2-\cos2\theta
\end{array} \right) \;\;.
\eea
The matrix $\Omega$ is orthogonal, therefore $S$ is also an orthogonal
matrix, (note that $(\Omega^T)^\alpha_{\;\;\;\beta}=\eta^{\alpha \alpha}
\eta_{\beta \beta} \Omega^\beta_{\;\;\;\alpha}$ with $\eta_{\mu \nu}=
diag(-1,1,...,1)$).

The boundary state equations (9)-(14) and (19) have the following solution
\bea  
\mid B_x , \tau_0   \rangle 
&=& \frac{T}{2}
\sqrt{1-E^2} \;\exp \bigg{[} i\alpha'\tau_0 \bigg{(}
(-p^1_{op}\sin \theta+p^2_{op}\cos \theta)^2 
+\sum^{d-1} _{j=3}(p^j_{op})^2 \bigg{)} \bigg{]}
\nonumber\\
&~&\times [\prod^{d-1}_{j=3} \delta(x^j-y^j)] 
\delta[-(x_1-y_1)\sin \theta+(x_2-y_2)\cos \theta]  
\nonumber\\
&~& \times \sum_{p^0} \sum_{p^1} \sum_{p^2}(\mid p^0 \rangle \mid p^1 
\rangle \mid p^2 \rangle ) \prod^{d-1}_{j=3} \mid p^j_L = p^j_R = 0 \rangle
\nonumber\\
&~& \times \exp \bigg{[} -\sum^{\infty}_{m=1} \bigg{(} \frac{1}{m}
e^{4im\tau_0} \alpha^{\mu}_{-m} S_{\mu \nu}(\theta,E) \tilde{\alpha}
^{\nu}_{-m} \bigg{)} \bigg{]} \mid 0 \rangle \;\;,
\eea
where  $T=\frac{\sqrt{\pi}} {2^{(d-10)/4}}(4\pi^2 \alpha')^{(d-6)/4}$ 
is the tension of $D_1$-brane in $d$-dimension \cite{5}.  
The last line is the solution of equation (19) and other factors 
are the solutions of 
equations (9)-(14). The left and right components of the momentum states 
$\mid p^\alpha \rangle= \mid p^\alpha_L \rangle \mid p^\alpha_R \rangle$
that are appeared in this state have the following relations
\bea
p^0 = \frac{1}{\alpha'} E({\ell}_1 \cos \theta +{\ell}_2 \sin \theta)\;\;,
\eea
\bea
p^1 = \frac{1}{\alpha'}E{\ell}^0 \cos \theta \;\;,
\eea
\bea
p^2 = \frac{1}{\alpha'}E{\ell}^0 \sin \theta \;\;,
\eea
\bea
{\ell}_1 \sin \theta = {\ell}_2 \cos \theta \;\;,
\eea
where $p^\mu = p^\mu _L + p^\mu _R $ and ${\ell}^\mu = 
\alpha'(p^\mu _L - p^\mu _R ) = N^\mu R^\mu$. We must consider equations
(23)-(26) in summing over $p^0 , p^1$ and $p^2$ in (22).
Energy of the closed string state depends on its winding numbers around
the $X^1$ and $X^2$ directions. According to the equations (24) and (25) 
for compact time, 
closed string state has non-zero momentum along the directions $X^1$
and $X^2$. Therefore its momentum numbers, along these directions 
(i.e. $M_1$ and $M_2$ ) are proportional to its winding number, around
the time direction.
Closed string can wind around the directions $X^1$
and $X^2$ if angle $\theta$ and radii of compactification $R_1$ and $R_2$
are such that the quantity $\frac{R_1 \sin \theta}{R_2 \cos \theta}$
is rational, otherwise $N_1=N_2=0$, i.e. closed string has no winding
around the $X^1$ and $X^2$, in this case its energy also is zero.

The ghost part of the boundary state is independent of $E$ and angle
$\theta$, it is
\bea
  \mid B_{gh}, \tau_{0} \rangle = \exp \bigg{[} 
  \sum_{m=1}^{\infty}{e^{4im\tau_0}}
  (c_{-m}{\tilde{b}}_{-m}-b_{-m}{\tilde{c}}_{-m})\bigg{]}
  \frac{c_0+ \tilde{c}_0}{2} \mid q=1 \rangle \mid \tilde{q} =1 \rangle\;\;.
\eea
{\bf Interaction}

  Now we can calculate the overlap of the two boundary states to obtain the  
  interaction amplitude of non-intersecting
  angled $m_1$ and $m_{1'}$ branes. Let
  $m_{1'}$ also be parallel to 
  the $X^1X^2$-plane with electric field $E'$ on it, 
  therefore boundary state that describes it, is given by equations
  (22)-(26) with
  the change $E \rightarrow E'$ , $y \rightarrow y' $ 
  and $\theta \rightarrow \theta'$. 
  These mixed branes simply
  interact via exchange of the closed strings, so that the amplitude is given
  by
\bea 
{\cal{A}} = \langle B_{1'},\;\tau_0 =0 \mid D 
\mid B_1,\;\tau_0 =0 \rangle \;\;,
\eea
where ``$D$'' is the closed string propagator.
In this amplitude we must use the total boundary state, 
i.e. $\mid B , \tau_0  \rangle = \mid B_x , \tau_0  \rangle 
\mid B_{gh} , \tau_0  \rangle$. Here we only give the final result;
\bea
{\cal{A}}_{bos} &=& \frac{T^2\alpha' L}{4(2\pi)^{d-2} \mid \sin \phi \mid}
\sqrt{(1-E^2)(1-E'^2)} 
\int_0^{\infty} dt \bigg{\{} e^{4at} 
\bigg( \sqrt{\frac{\pi}{\alpha't}} \; \bigg)^{d_{j_n}} 
\nonumber\\
&~&\times e^{ -\frac{1}{4\alpha't}\sum_{j_n}(y'^{j_n}-y^{j_n})^2 }
\prod_{j_c}\Theta_3 \bigg( \frac{y'^{j_c} 
- y^{j_c} }{2\pi R_{j_c}} \mid 
\frac{i\alpha't}{\pi (R_{j_c})^2}\bigg)
\nonumber\\
&~&\times \prod_{n=1}^\infty \bigg{[} (1-e^{-4nt})^{5-d}
[ det(1-\Omega' \Omega^T e^{-4nt})]^{-1} \bigg{]}
\Theta_3 (\nu \mid \tau) \bigg{\}} \;\;,
\eea
where $L=2\pi R_0$ is time length, $\phi = \theta-\theta'$. Also 
$\nu$ and $\tau$ have definitions,
\bea
&~&\nu = \frac{R_0}{2\pi \alpha' \sin \phi} \bigg{(} (E-E' \cos \phi)
\bar{y}'_2 + (E'-E \cos \phi) \bar{y}_2 \bigg{)} \;\;,
\nonumber\\
&~&\;\;\;\;\;\;\tau = \frac{itR^2_0}{\pi \alpha'} \bigg{(} \frac{E^2 +E'^2 
-2EE' \cos \phi}{\sin^2 \phi} -1 \bigg{)} \;\;,
\eea
and the determinant is
\bea
det(1- \Omega' \Omega^T e^{-4nt})= (1-e^{-4nt}) \bigg{[} 1- 2 e^{-4nt}
\bigg{(} -1 + \frac{2(EE'-\cos \phi)^2}{(1-E^2)(1-E'^2)} \bigg{)}
+e^{-8nt} \bigg{]} \;\;.
\eea
The set $\{{\bar{y}}_2,y_3,...,y_{d-1}\}$ 
shows the position of $m_{1}$-brane, with
$\bar{y}_2 = -y_1 \sin \theta + y_2 \cos \theta $ and $y_1 \cos \theta +y_2
\sin \theta =0$, similarly  
for $\bar{y}'_2$. The sets $\{j_n\}$
and $\{j_c\}$ are non-compact and compact part of $\{j\}$ , and 
$d_{j_n}$ is dimension of $\{X^{j_n}\}$. Because of the electric fields, this
amplitude is not symmetric under the exchange $\phi \leftrightarrow 
\pi-\phi$, on the other hand for angled mixed branes $\phi$ and $\pi-\phi$
are two different configurations.
From (30) we see that electric
fields and compactification of time cause $\bar{y}'_2$ and $\bar{y}_2$
to appear in the interaction. For non-compact time these disappear from the
interaction, i.e. $\Theta_3 (\nu \mid \tau )=1$ , as expected. The amplitude
(29) is symmetric with respect to the $m_1$ and $m_{1'}$ branes, 
\bea 
{\cal{A}}(E,E',y_1,y'_1, y_2,y'_2,y_j,y'_j,\theta,\theta')
= {\cal{A}}^*(E',E,y'_1,y_1,y'_2,y_2,y'_j,y_j,\theta',\theta) 
\eea
for complex conjugation see (28). It is also independent of the 
compactification of directions $X^1$ and $X^2$.
 For non-compact spacetime, remove all
factors $\Theta_3$ from (29) and change $j_n \rightarrow j$ therefore
$d_{j_n} \rightarrow d-3 $, in this case
interaction depends on the minimal distance between the branes, 
that is $\sum_j (y'^j -y^j )^2 $. When all directions are compact,
the factors containing $j_n$ disappear, in this case 
$j_c \in \{3,4,...,d-1\}$.
%%%%%%%%%%%%%%%%%%%%%%%%%%%%%%%%%%%%%%%%%%%%%%%%%%%%%%%%%%%%%%%%%%%%%%%%%%%%
\section{Interaction in superstring theory}

{\bf Boundary state}

 The boundary conditions on the fermionic 
 degrees of freedom should be imposed on 
both $R \otimes R$ and $NS \otimes NS$ sectors. World sheet supersymmetry
requires the two sectors to satisfy the boundary conditions,
\bea
\bigg{(}(\psi^0 -i \eta \tilde{\psi}^0) + E
( \psi^1 + i \eta \tilde{\psi}^1 )
\bigg{)}_{\tau_0} \mid B_{\psi} , \eta , \tau_0 \rangle = 0 \;\;,
\eea
\bea
\bigg{(}(\psi^1 -i \eta \tilde{\psi}^1) +E( \psi^0 + i \eta \tilde{\psi}^0 )
\bigg{)}_{\tau_0} \mid B_{\psi} , \eta , \tau_0 \rangle = 0 \;\;,
\eea
\bea
(\psi^i +i \eta \tilde{\psi}^i) 
_{\tau_0} \mid B_{\psi} , \eta , \tau_0 \rangle = 0 \;\;,
\eea
for the $m_1$-brane along $X^1$-direction \cite{8}. 
The parameter $\eta = \pm 1$
is introduced for the GSO projection. This state preserves half of the 
world sheet supersymmetry. For the rotated $m_1$-brane that makes angle
$\theta$ with $X^1$-direction, we must rotate $\psi^1$ and $\psi^2$,
therefore we find the following boundary states,
\bea
\mid B_{\psi} , \eta , \tau_0 \rangle _{NS} = \exp \bigg{[} i \eta
\sum^{\infty}_{r=1/2} \bigg{(} e^{4ir\tau_0} b^{\mu}_{-r} S_{\mu \nu}
(\theta,E) \tilde{b}^{\nu}_{-r} \bigg{)} \bigg{]} \mid 0 \rangle \;\;,
\eea
for the NS-NS sector, and
\bea
\mid B_{\psi} , \eta , \tau_0 \rangle _R = 
\frac{1}{\sqrt{1-E^2}}\; \exp \bigg{[} i \eta
\sum^{\infty}_{m=1} \bigg{(} e^{4im\tau_0} d^{\mu}_{-m} S_{\mu \nu}
(\theta,E) \tilde{d}^{\nu}_{-m} \bigg{)} \bigg{]} \mid B_{\psi} , \eta , 
\rangle^{(0)}_R \;\;,   
\eea
for the R-R sector. The origin of the factors $\sqrt{1-E^2}$ in (22) and
(37) is in the path integral with boundary action \cite{9}. Zero mode part
of the boundary state satisfies 
\bea
\bigg{(} d^{\mu}_0-i\eta S^{\mu}_{\;\;\;\nu}(\theta,E) \tilde{d}^{\nu}_0
\bigg{)} \mid B_{\psi} , \eta , \rangle^{(0)}_R =0\;\;.   
\eea
The vacuum for the fermionic zero modes $d^{\mu}_0$ and $\tilde{d}^{\mu}_0$
can be written as \cite{5}
\bea
\mid A \rangle \mid \tilde{B} \rangle = \lim_{z, \bar{z} \rightarrow 0}
S^A(z) \tilde{S}^B(\bar{z}) \mid 0 \rangle \;\;,
\eea
where $S^A$ and $\tilde{S}^B$ are the spin fields in the 32-dimensional
Majorana representation. We use a chiral representation for the $32
\times 32$ $\Gamma$-matrices of $SO(1,9)$ as in reference \cite{5},
therefore we consider solution of (38) of the form
\bea
\mid B_{\psi} , \eta \rangle ^{(0)}_R =
{\cal{M}}^{(\eta)}_{AB} \mid A \rangle \mid
\tilde{B} \rangle \;\;\;,
\eea
therefore the $32 \times 32$ matrix ${\cal{M}}^{(\eta)}$ obeys the
following equation
\bea
(\Gamma^{\mu})^T {\cal{M}}^{(\eta)} -i \eta S^{\mu}_{\;\;\;\nu}(\theta,E)
\Gamma_{11} {\cal{M}}^{(\eta)} \Gamma^{\nu} = 0 \;\; .
\eea
This equation has the solution
\bea
{\cal{M}}^{(\eta)} = C \Gamma^0(-E\Gamma^0+\Gamma^1 \cos \theta+\Gamma^2
\sin \theta) \bigg{(} \frac{1+i\eta \Gamma_{11}}{1+i\eta} \bigg{)}\;\;\;,
\eea
where $C$ is the charge conjugation matrix.

The superghost part of the NS-NS sector boundary state in the $(-1 , -1)$
picture is 
\bea
\mid B_{sgh} , \eta , \tau_0 \rangle _{NS} = \exp \bigg{[} i \eta
\sum^{\infty}_{r=1/2} e^{4ir\tau_0} (\gamma_{-r}\tilde{\beta}_{-r}
-\beta_{-r}\tilde{\gamma}_{-r}) \bigg{]} \mid P=-1 , 
\tilde{P}=-1 \rangle \;\;,
\eea
and for the R-R sector boundary state in the $(-1/2 , -3/2)$ picture is
\bea
\mid B_{sgh} , \eta , \tau_0 \rangle _{R} = \exp \bigg{[} i \eta
\sum^{\infty}_{m=1} e^{4im\tau_0} (\gamma_{-m}\tilde{\beta}_{-m}
-\beta_{-m}\tilde{\gamma}_{-m}) +i\eta \gamma_0 \tilde{\beta}_0
\bigg{]} \mid P=-1/2 , \tilde{P}=-3/2 \rangle \;\;,
\eea
where the superghost vacuum is annihilated by $\beta_0$ and 
$\tilde{\gamma}_0$ \cite{10}.

For both the NS-NS and the 
R-R sectors the complete boundary state can be written
as the following product
\bea
\mid B , \eta , \tau_0 \rangle _{R,NS}=
\mid B_{x} , \tau_0 \rangle \mid B_{gh} , \tau_0 \rangle 
\mid B_{\psi} , \eta , \tau_0 \rangle _{R,NS}
\mid B_{sgh} , \eta , \tau_0 \rangle _{R,NS}\;\;\;.
\eea
{\bf Interaction}

For calculation of the interaction amplitude, we must use the GSO projected
boundary states
\bea
\mid B , \tau_0 \rangle = \frac{1}{2}
(\mid B ,+, \tau_0 \rangle \mp \mid B , -, \tau_0 \rangle ) \;\;,
\eea
the minus sign is for the NS-NS and plus sign for the R-R sector.
In each sector the 
amplitude is given by (28) in which boundary states must be
replaced from (46). Finally the 
total amplitude, ${\cal{A}} = {\cal{A}}_{NS-NS}+ {\cal{A}}_{R-R}$, becomes
\bea
{\cal{A}} &=& \frac{T^2 \alpha' L}{8(2\pi)^8 \mid \sin \phi \mid}
\int_0^{\infty} dt \bigg{\{} 
\bigg( \sqrt{\frac{\pi}{\alpha't}} \; \bigg)^{d_{j_n}} 
e^{ -\frac{1}{4\alpha't}\sum_{j_n}(y'^{j_n}-y^{j_n})^2 }
\nonumber\\
&~& \times \Theta_3(\nu \mid \tau) 
\prod_{j_c}\Theta_3 \bigg( \frac{y'^{j_c} 
- y^{j_c} }{2\pi R_{j_c}} \mid 
\frac{i\alpha't}{\pi (R_{j_c})^2}\bigg) \bigg{(} \bigg{(} 
\sqrt{(1-E^2)(1-E'^2)}  
\nonumber\\
&~&\times \frac{1}{q} \bigg{[}\bigg{[}
\prod_{n=1}^\infty \bigg{[} \bigg{(} \frac{1+q^{2n-1}}{1-q^{2n}} \bigg{)}^5
\frac{\det(1+\Omega'\Omega^Tq^{2n-1})}{\det(1-\Omega'\Omega^Tq^{2n})}\bigg{]}
\nonumber\\
&~&-\prod_{n=1}^\infty \bigg{[} \bigg{(} \frac{1-q^{2n-1}}{1-q^{2n}} 
\bigg{)}^5 
\frac{\det(1-\Omega'\Omega^Tq^{2n-1})}{\det(1-\Omega'\Omega^Tq^{2n})}\bigg{]}
\;\;\bigg{]} \bigg{]}
\nonumber\\
&~&-16(\cos \phi-EE')\prod_{n=1}^\infty \bigg{[} \bigg{(} 
\frac{1+q^{2n}}{1-q^{2n}} \bigg{)}^5 
\frac{\det(1+\Omega' \Omega^Tq^{2n})}{\det(1-\Omega' \Omega^Tq^{2n})}
\bigg{]} \bigg{)} \bigg{)} \bigg{\}} \;\;\;.
\eea
where $q = e^{-2t}$. The last line comes from the R-R sector. The factor  
$(\cos \phi - EE')$ is contribution of the fermionic zero modes, according
to the sign and value of this factor, R-R interaction is repulsive,
attractive or zero. The other two terms come from the NS-NS sector.
Note that determinants in the denominators come from the world sheet
bosons and in the numerators from the fermions. These determinants have 
the expansion like (31), that $e^{-4nt}$ should be changed to $\pm q^{2n}$
and $\pm q^{2n-1}$ . This amplitude is symmetric with respect to $m_1$ and
$m_{1'}$ branes. 
Again for non-compact spacetime, in (47) remove all factors $\Theta_3$
and change $j_n \rightarrow j$ . 
%%%%%%%%%%%%%%%%%%%%%%%%%%%%%%%%%%%%%%%%%%%%%%%%%%%%%%%%%%%%%%%%%%%%%%%%%%%%%
\section{Interaction due to the massless states}

Now from the interaction amplitude (47), we extract the contributions of the 
NS-NS and the R-R sectors
massless states, to see how distant branes interact. Therefore 
we have the following limits \cite{6,8},
\bea
&~&\lim_{q \rightarrow 0} \frac{1}{q}\bigg{\{}
\prod_{n=1}^\infty \bigg{[} \bigg{(} \frac{1+q^{2n-1}}{1-q^{2n}} \bigg{)}^5
\frac{\det(1+\Omega' \Omega^Tq^{2n-1})}
{\det(1-\Omega' \Omega^Tq^{2n})}\bigg{]}
\nonumber\\
&~&-\prod_{n=1}^\infty \bigg{[} \bigg{(} 
\frac{1-q^{2n-1}}{1-q^{2n}} \bigg{)}^5 
\frac{\det(1-\Omega' \Omega^Tq^{2n-1})}
{\det(1-\Omega' \Omega^Tq^{2n})}\bigg{]}
\bigg{\}}
\nonumber\\
&~&=2[Tr(\Omega' \Omega^T)+5]
\nonumber\\
&~&=\frac{8}{(1-E^2)(1-E'^2)} 
\bigg{(} 1+\cos^2 \phi+2E^2 E'^2 -(E^2 + E'^2 
+2EE'\cos \phi) \bigg{)}
\eea
for the NS-NS sector, and 
\bea
\lim_{q \rightarrow 0} 
\prod_{n=1}^\infty \bigg{[} \bigg{(} \frac{1+q^{2n}}{1-q^{2n}} \bigg{)}^5
\frac{\det(1+\Omega' \Omega^Tq^{2n})}
{\det(1-\Omega' \Omega^Tq^{2n})}\bigg{]}=1
\eea
for the R-R sector. 

In the NS-NS sector, exchange of the massless states has the amplitude
\bea
{\cal{A}}^{NS-NS}_0 = \frac{T^2 \alpha' L}{(2\pi)^8 \mid \sin \phi \mid}
\frac{1+\cos^2 \phi+2E^2 E'^2 -(E^2 + E'^2 +2EE'\cos \phi)}{\sqrt{(1-E^2)
(1-E'^2)}}\;G\;\;\;,
\eea
\bea
G \equiv \int_0^{\infty} dt \bigg{\{} 
\bigg( \sqrt{\frac{\pi}{\alpha't}} \; \bigg)^{d_{j_n}} 
e^{ -\frac{1}{4\alpha't}\sum_{j_n}(y'^{j_n}-y^{j_n})^2 }
\Theta_3(\nu \mid \tau) 
\prod_{j_c}\Theta_3 \bigg( \frac{y'^{j_c} 
- y^{j_c} }{2\pi R_{j_c}} \mid 
\frac{i\alpha't}{\pi (R_{j_c})^2} \bigg{)} \bigg{\}}\;. 
\eea
According to the (50), the terms $(1+\cos^2 \phi +2E^2E'^2)$ have attractive
and $-(E^2+E'^2)$ have repulsive effects, the term $-EE' \cos \phi$ 
can have attractive or repulsive effect due to the signs of $E, E'$ and $\cos
\phi$. But, sum of all these terms is always positive, therefore exchange
of the massless states of the NS-NS sector produces attractive force
between these branes. 
In the $R-R$ sector, massless states have the following contribution
on the interaction
\bea
{\cal{A}}^{(R-R)}_0 =  \frac{T^2 \alpha' L}{(2\pi)^8
\mid \sin \phi \mid}[-2(\cos \phi -EE')] \;G\;\;,
\eea
according to the factor $(\cos \phi -EE')$,
this interaction can be attractive, repulsive or zero.
Therefore distant branes have interaction amplitude
${\cal{A}}_0 ={\cal{A}}^{(NS-NS)}_0+ {\cal{A}}^{(R-R)}_0$ . This is 
proportional to the factor $( \cos \phi -\cos \phi_0)^2$ 
, where
\bea
\cos \phi_0 = EE'+\sqrt{(1-E^2)(1-E'^2)}\;\;\;,  
\eea
so at $\phi=\phi_0$, attractive force of the NS-NS sector cancels the
repulsive force of the R-R sector.

For non-compact spacetime, the function $G$ is proportional to the Green's
function in seven dimensional space, i.e.
\bea
G_{(nc)} =\frac{1}{\alpha'}(2\pi)^7 G_7 (Y^2)
\eea
where $Y^2$ is minimal distance between 
the branes. In this space if $E=E'=0$, then massless
states amplitude ${\cal{A}}_0$ reduces to the \cite{11} with $p=1$.

%%%%%%%%%%%%%%%%%%%%%%%%%%%%%%%%%%%%%%%%%%%%%%%%%%%%%%%%%%%%%%%%%%%%%%%%%%%%
\section{Conclusion}

We explicitly determined the boundary state for both the NS-NS and R-R
sectors of superstring theory, corresponding to a mixed brane parallel to the
$X^1X^2$-plane. Energy of a closed string state emitted from this brane
depends on its winding numbers around $X^1$ and $X^2$ directions, radii
of compactification of these directions and back-ground internal 
electric field. Also,
for compact time closed string state can have non-zero momentum along the
brane. 

Compactification of time and non-zero electric fields 
imply that interaction should
depend on the positions $\bar{y}'_2$ and $\bar{y}_2$ of the branes.  
Depending on the back-ground fields and angle 
between the branes, R-R interaction 
is repulsive, attractive, or zero. For both of the NS-NS and the R-R
sectors, we extracted contribution of the massless states on the interaction.
For non-compact spacetime, these are proportional to the Green's function
of seven dimensional space. 

The formalism can be extended to include mixed 
branes with arbitrary dimensions
$p_1$ and $p_2$ and more than one angle.

{\bf acknowledgement}

The author would like to thank H.Arfaei and M.M. Sheikh-Jabbari
for fruitful discussions.

%%%%%%%%%%%%%%%%%%%%%%%%%%%%%%%%%%%%%%%%%%%%%%%%%%%%%%%%%%%%%%%%%%%%%%%%%%%%


\begin{thebibliography}{99}
\bibitem{1}
M. Frau, I. Pesando, S. Sciuto, A. Lerda and R. Russo,
Phys. Lett.{\bf B400}(1997)52, hep-th/9702037.
\bibitem{2}
M.B. Green, P. Wai, Nucl. Phys. {\bf B431} (1994)131; M. Li, Nucl. Phys.
{\bf B460} (1996)351, hep-th/9510161; C. Schmidhuber, Nucl. Phys.
{\bf B467}(1996)146, hep-th/9601003.
\bibitem{3}
M. Billo, P. Di Vecchia and D. Cangemi, Phys. Lett.{\bf B400}(1997)63, 
hep-th/9701190.
\bibitem{4}
F. Hussain, R. Iengo, C. Nunez, Nucl. Phys.{\bf B497}(1997)
205, hep-th/9701143.
\bibitem{5}
P. Di Vecchia, M. Frau, I. Pesando, S. Sciuto, A. Lerda and R. Russo,
 Nucl. Phys.{\bf B507}(1997)259, hep-th/9707068.
\bibitem{6}
H. Arfaei and D. Kamani, Phys. Lett.{\bf B452}(1999)54.
\bibitem{7}
M. Billo, P. Di Vecchia M. Frau, A. Lerda, I. Pesando, R. Russo. 
S. Sciuto, Nucl. Phys.{\bf B526}(1998)199, 
hep-th/9802088.
\bibitem{8}
H.Arfaei and D.Kamani ``Mixed Branes Interaction in Compact Spacetime''
to appear in Nucl. Phys.{\bf B}.
\bibitem{9}
C.G. Callan, C. Lovelace, C.R. Nappi and S.A. Yost, 
Nucl. Phys.{\bf B308} (1988)221-284.
\bibitem{10}
S.A. Yost, Nucl. Phys. {\bf B321}(1989)629-652.
\bibitem{11}
H. Arfaei and M.M. Sheikh-Jabbari, Phys. Lett. {\bf B394}(1997)288.


\end{thebibliography}
\end{document}